\documentclass{appolb}
\usepackage{graphicx}
\usepackage{epsfig}
\begin{document}
\title{Recent direct reaction experimental studies with radioactive tin beams
\thanks{Presented at the Zakopane Conference on Nuclear Physics ``Extremes of the Nuclear Landscape", August 31 - September 7, 2014, Zakopane, Poland.} }
\author{ K.L.~Jones$^{a}$, S.~Ahn$^{a,b}$, J.M.~Allmond$^{c}$, A.~Ayres$^{a}$, D.W.~Bardayan$^{c,d}$, T.~Baugher$^{b,e}$, D.~Bazin$^{b}$, J.S.~Berryman$^{b}$, A.~Bey$^{a}$, C.~Bingham$^{a,c}$,  L.~Cartegni$^{a}$, G.~Cerizza$^{a}$, K.Y.~Chae$^{c,f}$,  J.A.~Cizewski$^{e}$, A.~Gade$^{b}$, A.~Galindo-Uribarri$^{c}$, R.F.~Garcia-Ruiz$^{g}$, R.~Grzywacz$^{a,c}$,   M.E.~Howard$^{e}$, R.L.~Kozub$^{h}$, J.F.~Liang$^{c}$, B.~Manning$^{e}$, M.~Mato\v{s}$^{i,j}$, S.~McDaniel$^b$, D.~Miller$^{a,k}$, C.D.~Nesaraja$^{c}$, P.D.~O'Malley$^{e,d}$, S.~Padgett$^{a,l}$, E.~Padilla-Rodal$^{m}$, S.D.~Pain$^{c}$, S.T.~Pittman$^{c,i}$, D.C.~Radford$^{c}$, A.~Ratkiewicz$^{b,e}$, K.T.~Schmitt$^{c}$, A.~Shore$^{b}$, M.S.~Smith$^c$, D.W.~Stracener$^c$, S.R.~Stroberg$^{b}$, J.~Tostevin$^{n}$, R.L.~Varner$^{c}$, D.~Weisshaar$^b$, K.~Wimmer$^{b,o}$, R.~Winkler$^b$
\address{ 
$^a$Department of Physics and Astronomy, University of Tennessee, Knoxville, TN 37996, USA \\
$^b$National Superconducting Cyclotron Laboratory and Department of Physics and Astronomy, Michigan State University, East Lansing, MI 48824, USA.\\
$^c$Physics Division, Oak Ridge National Laboratory, Oak Ridge, TN 37831, USA. \\
$^d$ Department of Physics, University of Notre Dame, Notre Dame, IN 46556, USA. \\
$^e$Department of Physics and Astronomy, Rutgers University, New Brunswick, NJ 08903, USA. \\
$^f$Department of Physics, Sungkyunkwan University, Suwon 440-746, Korea \\
$^g$ Instituut voor Kernen Stralingsfysica, KU Leuven, B-3001, Leuven, Belgium \\
$^h$  Department of Physics, Tennessee Technological University, Cookeville, TN 38505, USA \\
$^i$ Department of Physics and Astronomy, Louisiana State University, Baton Rouge, LA 70803, USA \\
$^j$ International Atomic Energy Agency, Division of Physical and Chemical Sciences, P.O. Box 100, A-1400 Vienna, Austria \\
$^k$TRIUMF, Vancouver, BC, V6T 2A3C, Canada.\\
$^l$Lawrence Livermore National Laboratory, Livermore, CA 94550, USA.\\
$^m$ Instituto do Ciencias Nucleares, UNAM, AP 70-543, 04510 M\'{e}xico, D.F., M\'{e}xico \\
$^n$Department of Physics, University of Surrey, Guildford, Surrey GU2 7XH, UK. \\
$^o$Department of Physics, Central Michigan University, Mt Pleasant, MI 48858, USA.
}}
\maketitle

\begin{abstract}
Direct reaction techniques are powerful tools to study the single-particle nature of nuclei.  Performing direct reactions on short-lived nuclei requires radioactive ion beams produced either via fragmentation or the Isotope Separation OnLine (ISOL) method.  Some of the most interesting regions to study with direct reactions are close to the magic numbers where changes in shell structure can be tracked.  These changes can impact the final abundances of explosive nucleosynthesis.  The structure of the chain of tin isotopes is strongly influenced by the $Z=50$ proton shell closure, as well as the neutron shell closures lying in the neutron-rich, $N=82$, and neutron-deficient, $N=50$, regions.  Here we present two examples of direct reactions on exotic tin isotopes.  The first uses a one-neutron transfer reaction and a low-energy reaccelerated ISOL beam to study states in $^{131}$Sn from across the $N=82$ shell closure.  The second example utilizes a one-neutron knockout reaction on fragmentation beams of neutron-deficient $^{106,108}$Sn.  In both cases, measurements of $\gamma$ rays in coincidence with charged particles proved to be invaluable. \\

\end{abstract}


\section{Introduction}

Direct reactions have been used for decades to study the structure of atomic nuclei.  As more radioactive ion beams (RIBs) have become  available, the use of direct reactions on exotic nuclei has increased.  Here we discuss two types of direct reactions used with RIBs, transfer reactions, in particular neutron-adding transfer reactions, and nucleon knockout.  The former is best used with beams around the energy of the Coulomb barrier and up to a few 10's of MeV/A.  Reaccelerated Isotope Separation OnLine (ISOL) beams lend themselves to transfer reaction measurements, although lower-energy or degraded fragmentation beams have also been used.  Knockout reactions, however, require higher energy beams, typically around $50-250$~MeV/A.

\begin{figure}
\includegraphics[height=.4\textheight]{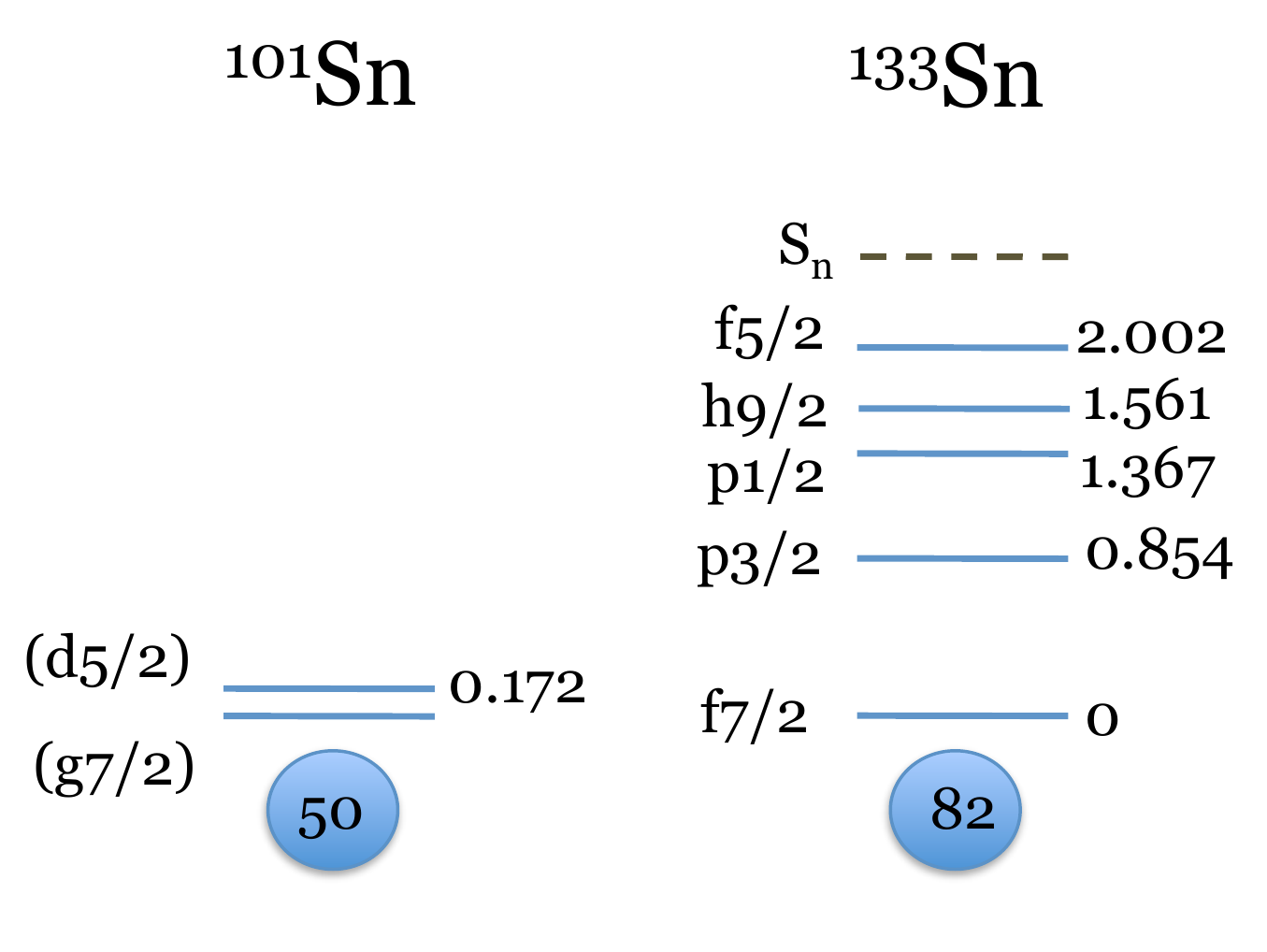}
\caption{
Schematic comparison of states measured in $^{101}$Sn and $^{133}$Sn.  Data taken from \cite{Sew07,Hof96,Jon10,Urb99}  Energies are shown in MeV.\label{tin_comp}}
\end{figure}

Two examples are given here, both using direct reactions on exotic tin beams.  Tin has the highest number of stable isotopes of any element, reflecting the magic nature of $Z=50$.  Structure studies of neutron-rich and neutron-deficient tin isotopes can probe shell evolution across a long chain of isotopes spanning two neutron shell closures.  On the neutron-deficient side, $^{101}$Sn, which is one neutron beyond presumably doubly magic $^{100}$Sn, has an excited state just 172 keV above its ground state.  Currently, the spin-parity assignments of these near degenerate states have not been confirmed experimentally.  However, there is evidence, combining information from double-$\alpha$ decay and state-of-the-art shell model calculations, that the ground and first excited state J$^{\pi}$ assignments are reversed in $^{101}$Sn compared to $^{103}$Sn \cite{Dar10} .  No other states in $^{101}$Sn have been observed to date.  The excited states in $^{133}$Sn, one neutron past doubly magic $^{132}$Sn, are well spaced up to the separation energy as shown in figure \ref{tin_comp}.

The first example presented here is a one-neutron transfer experiment on a reaccelerated ISOL beam of $^{130}$Sn.  The experiment was performed at the Holifield Radioactive Ion Beam Facility (HRIBF) \cite{Bee11}  at Oak Ridge National Laboratory, during the last period that it ran as a user facility, in Spring 2012.  Previously, an inverse kinematics (d,p) reaction had been performed, also at the HRIBF \cite{Koz12}, measuring proton ejectiles from the final state.  In reference \cite{Koz12}, four states were observed for the first time, and were shown to have a largely single-particle nature.  However, resolution in the (d,p) experiment was limited, leading to large uncertainties ($\pm 50$~keV) in the energies of these states.  The transfer experiment described here used a heavy-ion induced one-neutron-adding reaction, and critically, incorporated $\gamma$-ray detection in coincidence with charged particles.

The second example is the one-neutron knockout reaction on beams of $^{108}$Sn and $^{106}$Sn.  Here the experiment utilized fragmentation beams from the National Superconducting Cyclotron Laboratory (NSCL) at Michigan State University.  

The goal of this paper is to show the contrasting direct reaction techniques for measuring the structure of these different exotic tin isotopes using low-energy ISOL and high-energy fragmentation beams.

\section{$^{131}$Sn studied with the ($^9$Be, $^8$Be $\gamma$) reaction}
A series of detailed transfer reaction studies of nuclei around $^{132}$Sn \cite{Koz12,Jon11, All14, All12} and $N=50$ above $Z=28$ \cite{Tho05, Tho07} were performed at the HRIBF.  Inverse kinematics (d,p) experiments on beams of $^{130}$Sn and $^{132}$Sn resulted in the population of excited states of $^{131}$Sn from across the N = 82 shell closure with properties similar to those seen at low excitations in $^{133}$Sn.  These states  were further studied using the ($^9$Be,$^8$Be $\gamma$) reaction on beams of $^{130}$Sn \cite{Bey14} and $^{132}$Sn \cite{All14}. The focus here is on the $^{130}$Sn beam results.    The tin nuclei, along with a large number of neighboring nuclei, were produced by proton-induced fission on a thick natural uranium carbide target.  A purified beam of tin ions was produced by adding H$_2$S gas to the target system to form tin sulfide molecules, which were then ionized in the hot plasma ion source.  This technique greatly suppresses the isobaric contamination in the negative-ion beam delivered to the post-accelerator, leading to a beam that was greater than 99\% pure $^{130}$Sn.The beam,with an intensity of around 10$^5$ particles per second, was accelerated to a total energy of 520~MeV and  impinged on a 2~mg/cm$^2$ natural beryllium target . 

\begin{figure}
\includegraphics[trim=50 50 0 0,clip,height=.4\textheight]{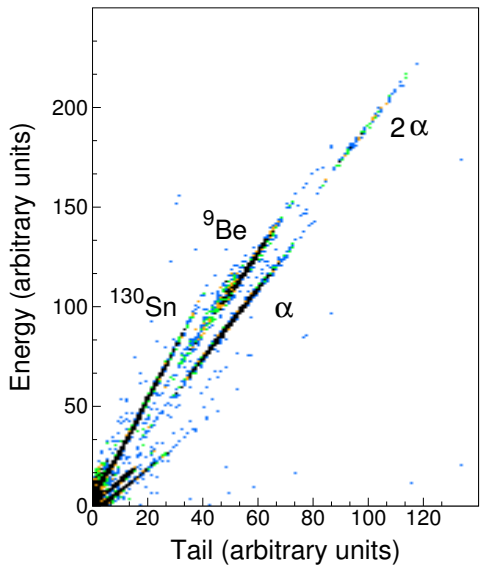}
\caption{
Particle identification plot from the $^{130}$Sn($^9$Be,$^8$Be $\gamma$)$^{131}$Sn inverse kinematics reaction as measured in one of the 54 crystals of the BareBall.\label{PID-8Be}}
\end{figure}

 Charged particles from the reaction were measured in the BareBall array of 54 CsI detectors \cite{Gal10}.  Pulse shape discrimination was used to give clear particle identification of light reaction products.  The total energy of each CsI signal was measured, as was the slow component, referred to as the tail, as shown in Figure \ref{PID-8Be}.  Separate groups for $\alpha$ particles, $^9$Be ions, and scattered $^{130}$Sn beam particles can be clearly resolved.  The $^8$Be ions produced in the reaction break up into two closely correlated $\alpha$ particles.  These $\alpha$ particles may strike the same crystal giving a very clean tag of the reaction, the 2$\alpha$ group marked on Figure \ref{PID-8Be}.  Otherwise the two $\alpha$ particles may be detected in two neighboring crystals, which again leads to a clean tag by inspecting the relevant multiplicity-2 events.  In the case where only one of the two $\alpha$ particles was detected, the event would be placed in the same place in Figure \ref{PID-8Be} as events where only one $\alpha$ particle was emitted from the reaction, but unlike the previous case the multiplicity would be 1.  This leads to a much less clean tag of the channel of interest.  The $\gamma$-ray data could then be gated on these different charged-particle conditions.  Only the 2$\alpha$ and multiplicity 2 charged-particle gates were used in the analysis of the $\gamma$-ray energies.  

\begin{figure}
\includegraphics[height=.4\textheight]{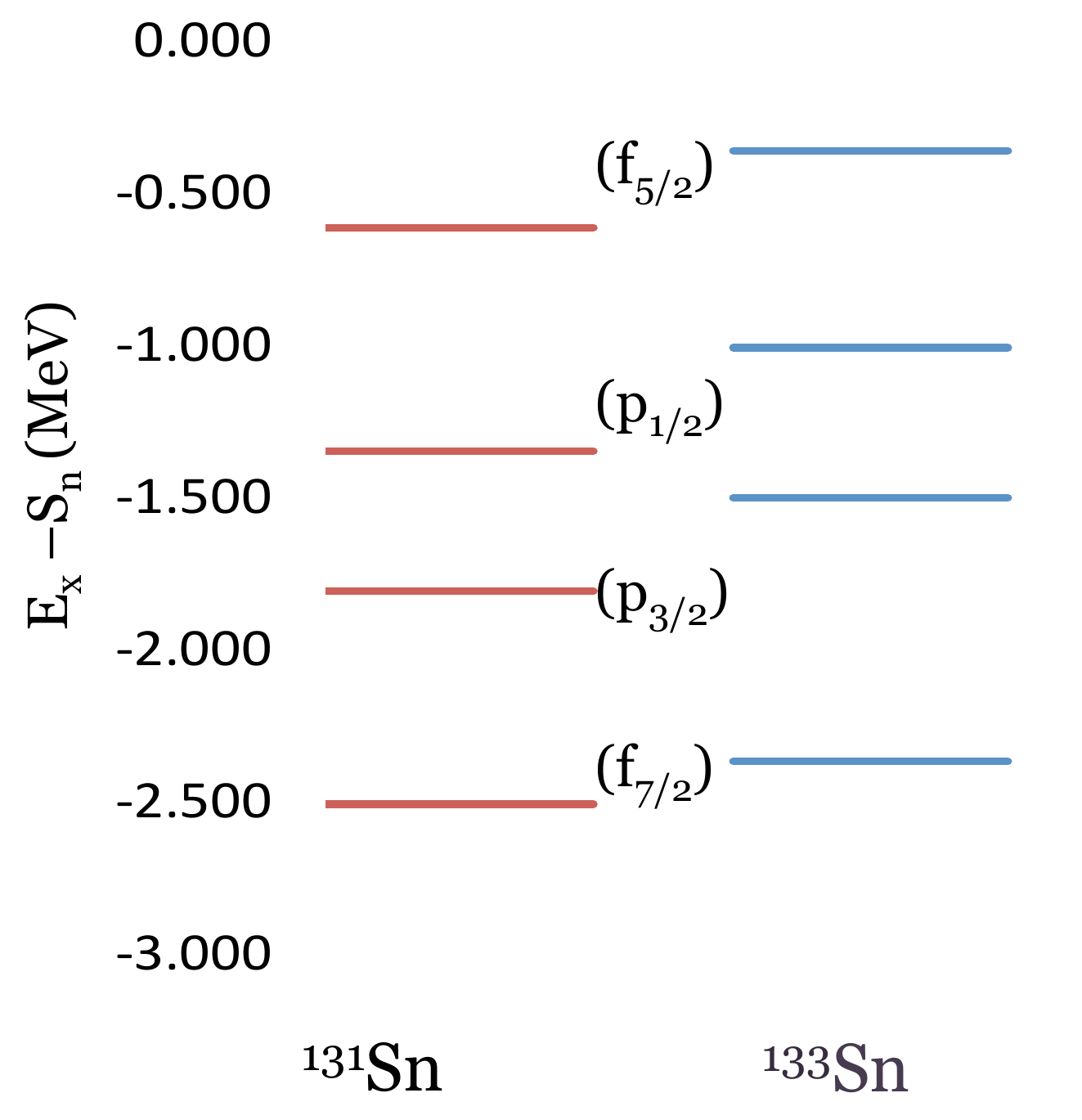}
\caption{
Comparison of the energy below the neutron separation energy for states in $^{131}$Sn and $^{133}$Sn observed in inverse kinematics neutron-transfer measurements \cite{Koz12,Jon11}.  Data also taken from \cite{Hof96,Urb99,All14}\label{133_comp}}
\end{figure}

The precise locations of the states observed in reference \cite{Koz12} have been determined from the $\gamma$ rays measured in the CLARION HPGe array \cite{Gro00}. The four states measured for the first time in the (d,p) experiment, presumably with the structure of two holes below $N=82$ coupled to the neutron f$_{7/2}$, p$_{3/2}$, p$_{1/2}$, and f$_{5/2}$ single-particle states from above the $N=82$ shell closure (1p-2h), were observed here via their $\gamma$ decays. Excitation energies of these 1p-2h states were found to agree with those measured in the (d,p) experiment to within 1$\sigma$ in three cases, and with a larger discrepancy of approximately 2$\sigma$ for the (7/2$^-$) state. As these single-particle like states are from across the shell closure at $N=82$ they have high excitation energies ($E_x=2.659$ to $4.554$~MeV) in $^{131}$Sn, whereas in $^{133}$Sn the 7/2$^-$ state, for example, is the ground state.  Therefore, in order to make a meaningful comparison of the energies of these states in $^{131}$Sn and $^{133}$Sn it is necessary to subtract the excitation energy from the neutron separation energy, giving the binding energy of each state.  Figure \ref{133_comp} shows a remarkable similarity in the binding energies of these four single-particle states in $^{133}$Sn and the corresponding 1p-2h states in $^{131}$Sn.

Known transitions from high-spin states in $^{131}$Sn \cite{Bha01}, including the 284- and 4273-keV transitions from the (19/2$^+$) state at 4558 keV,  were also observed following the single-neutron transfer reaction here.  In these reactions at energies close to the Coulomb barrier, states will be weakly populated if more than about four units of angular momentum need to be transferred.  Therefore, transfers on the 0$^+$ ground state of $^{130}$Sn to states with spins much above 9/2 are suppressed.  Other mechanisms, such as fusion evaporation, can be discounted owing to the clean 2$\alpha$ particle detection requirement.  However, the 7$^-$ isomer of $^{130}$Sn is a known component of this beam, making up approximately 7\% of the purified beam \cite{Str}.  Transferring three units of angular momentum, for example,  onto this metastable state can populate states up to spin 21/2.  This is the most reasonable explanation for how these states were populated in this reaction \cite{Bey14}.  Similar states were not observed in experiments using the ($^9$Be,$^8$Be $\gamma$) reaction on beams of $^{132}$Sn \cite{All14} and $^{134}$Te \cite{All12}, which do not have an isomeric component.

\section{$^{105,107}$Sn studied with one-neutron knockout reactions}
At the other end of the tin isotopic chain, the region around $^{100}$Sn is influenced by the N=Z=50 magic numbers, the N=Z line, and is also close to the proton drip-line, and the end of the rp-process. Despite this, there is little spectroscopic information available on light tin isotopes owing mostly to low production rates.  One important step in understanding the shell structure in this region is to measure the single-particle states, and study how they evolve with neutron number. Until recently, there were no firm J$^{\pi}$ assignments for odd-mass tin isotopes lighter than $^{109}$Sn. 

Knockout reactions can be used with very weak beams, down to just a few particles per second, and are sensitive to the $\ell$ value of the knocked-out neutron \cite{Han01,Han03,Gad08a}.  In order to make exclusive measurements, that is to separate the states populated in the residual nucleus, it is essential to measure $\gamma$ rays in coincidence with the recoiling charged particle.

An experiment was performed at the NSCL using 80~MeV/A (at the middle of the secondary target) fragmentation beams of $^{108,106}$Sn on a 47~mg/cm$^2$ $^9$Be target at rates of $3\times10^3$ and 1$\times10^2$ pps, respectively.  The beams were produced by impinging a 140~MeV/A $^{124}$Xe primary beam on a 240~mg/cm$^2$ $^9$Be target and tuning the A1900 \cite{Mor03} for the secondary beam of interest.  The experiment used high-efficiency CsI(Na) $\gamma$ ray detector array CAESAR \cite{Wei10}. The reaction residues were detected and identified with the S800 focal plane detector system \cite{Baz03}.  
\begin{figure}
\includegraphics[trim=50 50 0 100,clip,height=.45\textheight]{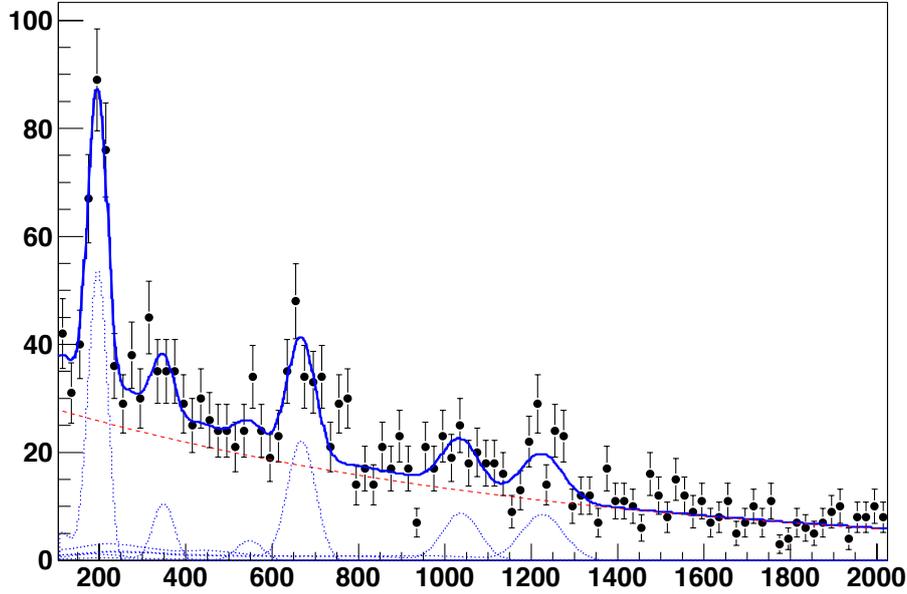}
\caption{
(Color online) Gamma-ray spectrum following the one-neutron knockout from $^{106}$Sn.  The solid, blue curve is a fit using the background shown by the dashed, red line.  The dotted line shows the GEANT simulation of the 200-keV gamma decay from the first excited state in $^{105}$Sn, and other higher-lying states populated via the knockout of deeper bound neutrons. \label{105_gamma}}
\end{figure}
 \begin{figure}
\includegraphics[trim=20 50 0 100,clip,height=.4\textheight]{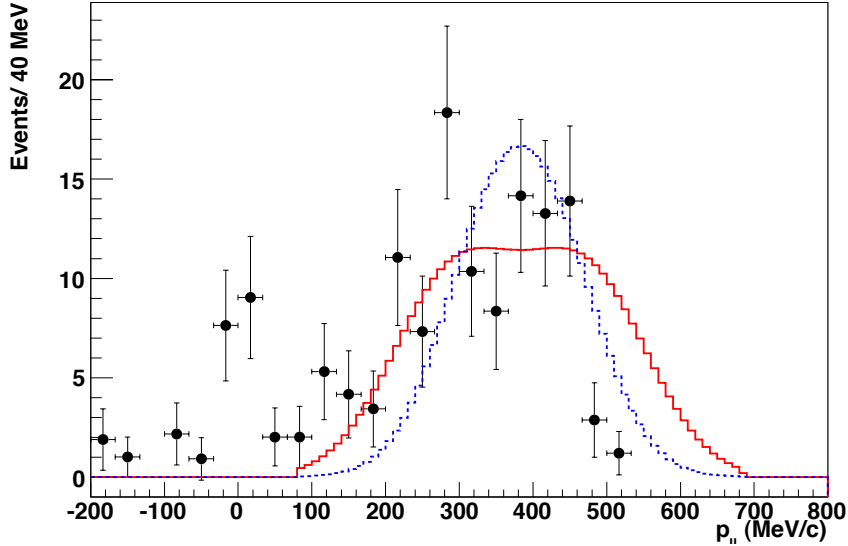}
\caption{
(Color online) Measured momentum distribution, gated on the 200-keV $\gamma$-ray from the first excited state in $^{105}$Sn, following the one-neutron knockout from $^{106}$Sn.  The red solid (blue dashed) line shows the Glauber model calculation assuming the knockout of a g$_{7/2}$ (d$_{5/2}$) neutron.\label{105_mom}}
\end{figure}
Here we show the analysis of the data from the $^{106}$Sn beam; the data from the $^{108}$Sn beam are the subject of a forthcoming publication \cite{Cer14}.  Data on the $^{106}$Sn beam were collected for 28~hours, leading to suboptimal statistics.  However, the decay of the first excited state of $^{105}$Sn at 200~keV can be seen clearly in Figure \ref{105_gamma}.   The dotted lines show the expected signal from the 200-keV decay from the first excited state in $^{105}$Sn, without background taken into account, and other peaks coming from higher-lying states in $^{105}$Sn populated through the knockout of neutrons from below $N=50$ in $^{106}$Sn.  \\
The position of the $^{105}$Sn fragment on the S800 focal plane was measured in cathode readout drift chambers. Combining this position information with the reconstructed trajectory through the spectrometer, using the {\sc cosy infinity} code \cite{Mak99}, allowed the full momentum vector to be reconstructed on an event-by-event basis.  The momentum distribution for the one-neutron knockout from $^{106}$Sn is difficult to interpret due to the low statistics.  Calculations using the Glauber model \cite{Ber04} assuming that the knocked-out neutron was g$_{7/2}$ in nature is shown by the solid, red line.  This gives a reasonable fit, considering the statistical uncertainties, except for the two highest momentum points.  The data are also compatible with the dashed, blue line, showing the calculation assuming knock-out of a d$_{5/2}$  neutron.  The ground state of $^{105}$Sn is expected to be 5/2$^+$ and the first excited state 7/2$^+$, both from extrapolation from heavier tin isotopes, and from shell model calculations.  The data are not of a sufficient quality to discriminate between a 7/2$^+$ and 5/2$^+$  assignment for the first excited state of $^{105}$Sn.

\section{Summary and Outlook}
Data from both transfer and knockout reactions can provide insight into the structure of exotic nuclei.  Where beams of at least 10$^4$ to 10$^5$ particles per second are available with energies near the Coulomb barrier, up to a few 10's of MeV/A, transfer reactions can be used to study the ground and excited states of the residual nucleus.  Knockout reactions can be used with much weaker beams, around a few particles per second, at higher energies with much thicker targets.  Here we presented preliminary results from two experiments utilizing beams of exotic tin isotopes.  In both the knockout and the transfer cases the measurement of $\gamma$ rays in coincidence with charged particles proved necessary to measure the energies of states, or to separate states that could not otherwise be resolved.  Additionally, the observation of previously known $\gamma$-ray transitions from high-spin states in $^{131}$Sn \cite{Bha01} led to the recognition of the first transfer reaction on an isomeric, rare ion beam.

Transfer reactions on radioactive ion beams with coincident measurement of $\gamma$ rays are being performed at various laboratories around the world.  Commonly, the requirement to cover a large solid angle calls for compact geometries, with particle detectors being placed close to the target, compromising the resolution in Q-value.  Combining ORRUBA, the Oak Ridge Rutgers University Barrel Array of position sensitive silicon strip detectors \cite{Pai07} and the Gammasphere \cite{Lee90} array of 110 Compton-suppressed HPGe detectors, into the Gammasphere-ORRUBA Dual Detectors for Experimental Structure Studies, GODDESS \cite{Rat13,Rat12, Pai14}, provides high $\gamma$-ray resolution and efficiency, without compromising measurements of the charged-particles.  This is possible because the full ORRUBA can fit in the Gammasphere target chamber.  Further plans include transfer reaction studies on $^{252}$Cf fission fragments at Argonne National Laboratory \cite{Sav08}.

Knockout measurements on light tin isotopes are planned to continue at the RIKEN Radioactive Ion Beam Factory including the critical measurement of the one-neutron knockout from $^{102}$Sn.  This measurement will be used to make J$^{\pi}$ assignments to the ground and first excited states in $^{101}$Sn.

Direct reactions across a broad energy regime will continue to be a vital tool at present and next-generation rare-isotope facilities.  The strengths of these techniques performed with beams of exotic nuclei have been proven by various groups over the last 30 years and instrumentation related to direct reaction measurements is being designed and built to match the capabilities of new facilities.

\section{Acknowledgements}
 This material is based upon work supported by the US Department of Energy, Office of Science,  Office of Nuclear Physics under contract numbers DE-FG02-96ER40983 (UT), DE-SC0001174(UT), DE-FG02-96ER40955 (TTU), DE-AC05-00OR22725 (ORNL), DE-FG02-96ER40978 (LSU), the U. S.  Department of Energy, National Nuclear Security Administration Stewardship Science Academic Alliance Program under contracts DE-FG52-08NA28552  and DE-NA0002132 (Rutgers) and by  the National Science Foundation under grant PHY-1067806 (Rutgers), PHY-1102511 and PHY-0722822 (NSCL), and the United Kingdom Science and Technology Facilities Council (STFC) under Grant No. ST/J000051/1. This research was conducted partly at the Oak Ridge National Laboratory Holifield Radioactive Ion Beam Facility, a DOE Office of Science User Facility. \\

\bibliographystyle{aip}
\bibliography{zak2014_Jones}
\end{document}